\documentclass[times,preprinttwo]{aastex62}
\usepackage{amsfonts,amsmath,amssymb,amsthm,epsfig,graphicx,float,tabularx,multirow,booktabs,gensymb}
\usepackage{booktabs,color,textcomp}
\usepackage{natbib}
\usepackage{rotating}
\usepackage{makecell}
\usepackage{multirow}

\setcellgapes{0pt}

\newcolumntype{S}{>{\centering\arraybackslash}m{0.05\linewidth}}
\newcolumntype{E}{>{\centering\arraybackslash}m{0.05\linewidth}}
\newcolumntype{M}{>{\centering\arraybackslash}m{0.1\linewidth}}
\newcolumntype{B}{>{\centering\arraybackslash}m{0.1\linewidth}}
\newcolumntype{N}{>{\centering\arraybackslash}m{0.0005\linewidth}}
\received{September 30th, 2019}
\revised{November 18th, 2019}
\accepted{November 21st, 2019}

\begin{document}

\title{First Resolved Scattered-Light Images of Four Debris Disks in Scorpius-Centaurus with the Gemini Planet Imager}

\correspondingauthor{Justin Hom}
\email{jrhom@asu.edu}

\author[0000-0001-9994-2142]{Justin Hom}
\affiliation{School of Earth and Space Exploration, Arizona State University, Tempe, AZ 85281, USA}

\author{Jennifer Patience}
\affiliation{School of Earth and Space Exploration, Arizona State University, Tempe, AZ 85281, USA}

\author[0000-0002-0792-3719]{Thomas M. Esposito}
\affiliation{Astronomy Department, University of California, Berkeley, CA 94720, USA}

\author[0000-0002-5092-6464]{Gaspard Duch\^ene}
\affiliation{Astronomy Department, University of California, Berkeley, CA 94720, USA}
\affiliation{Universit\'e Grenoble Alpes / CNRS, Institut de Plan\'etologie et d'Astrophysique de Grenoble, 38000 Grenoble, France}

\author{Kadin Worthen}
\affiliation{School of Earth and Space Exploration, Arizona State University, Tempe, AZ 85281, USA}

\author[0000-0002-6221-5360]{Paul Kalas}
\affiliation{Astronomy Department, University of California, Berkeley, CA 94720, USA}
\affiliation{SETI Institute, Carl Sagan Center, 189 Bernardo Ave,  Mountain View CA 94043, USA}
\affiliation{Institute of Astrophysics, FORTH, GR-71110 Heraklion, Greece}

\author{Hannah Jang-Condell}
\affiliation{Department of Physics \& Astronomy, University of Wyoming, Laramie, WY 82071, USA}

\author{Kezman Saboi}
\affiliation{School of Earth and Space Exploration, Arizona State University, Tempe, AZ 85281, USA}

\author[0000-0001-6364-2834]{Pauline Arriaga}
\affiliation{Department of Physics \& Astronomy, 430 Portola Plaza, University of California, Los Angeles, CA 90095, USA}

\author[0000-0002-9133-3091]{Johan Mazoyer}
\altaffiliation{NASA Sagan Fellow}
\affiliation{NASA Jet Propulsion Laboratory, California Institute of Technology, Pasadena, CA 91109, USA}

\author[0000-0002-9977-8255]{Schuyler Wolff}
\affiliation{Leiden Observatory, Leiden University, P.O. Box 9513, 2300 RA Leiden, The Netherlands}

\author[0000-0001-6205-9233]{Maxwell A. Millar-Blanchaer}
\altaffiliation{NASA Hubble Fellow}
\affiliation{NASA Jet Propulsion Laboratory, California Institute of Technology, Pasadena, CA 91109, USA}

\author[0000-0002-0176-8973]{Michael P. Fitzgerald}
\affiliation{Department of Physics \& Astronomy, 430 Portola Plaza, University of California, Los Angeles, CA 90095, USA}

\author[0000-0002-3191-8151]{Marshall D. Perrin}
\affiliation{Space Telescope Science Institute, Baltimore, MD 21218, USA}

\author[0000-0002-8382-0447]{Christine H. Chen}
\affiliation{Space Telescope Science Institute, Baltimore, MD 21218, USA}

\author[0000-0003-1212-7538]{Bruce Macintosh}
\affiliation{Kavli Institute for Particle Astrophysics and Cosmology, Stanford University, Stanford, CA 94305, USA}

\author[0000-0003-3017-9577]{Brenda C. Matthews}
\affiliation{National Research Council of Canada Herzberg, 5071 West Saanich Road, Victoria, BC V9E 2E7, Canada}

\author[0000-0003-0774-6502]{Jason J. Wang}
\affiliation{Astronomy Department, California Institute of Technology, Pasadena, CA 91126, USA}

\author{James R. Graham}
\affiliation{Astronomy Department, University of California, Berkeley, CA 94720, USA}

\author[0000-0001-7016-7277]{Franck Marchis}
\affiliation{SETI Institute, Carl Sagan Center, 189 Bernardo Avenue, Mountain View, CA 94043, USA}

\author{S. Mark Ammons}
\affiliation{Lawrence Livermore National Laboratory, 7000 East Ave., Livermore, CA 94550}

\author[0000-0002-5407-2806]{Vanessa P. Bailey}
\affiliation{NASA Jet Propulsion Laboratory, California Institute of Technology, Pasadena, CA 91109, USA}

\author[0000-0002-7129-3002]{Travis Barman}
\affiliation{Lunar and Planetary Laboratory, University of Arizona, Tucson AZ 85721 USA}

\author{Joanna Bulger}
\affiliation{Pan-STARRS Observatory, Institute for Astronomy, University of Hawai’i, 2680 Woodlawn Drive, Honolulu, HI 96822, USA}

\author[0000-0001-6305-7272]{Jeffrey K. Chilcote}
\affiliation{Kavli Institute for Particle Astrophysics and Cosmology, Stanford University, Stanford, CA 94305, USA}
\affiliation{Department of Physics, University of Notre Dame, 225 Nieuwland Science Hall, Notre Dame,
IN, 46556, USA}

\author[0000-0003-0156-3019]{Tara Cotten}
\affiliation{Physics and Astronomy, University of Georgia, 240 Physics, Athens, GA 30602, USA}

\author[0000-0002-4918-0247]{Robert J. De Rosa}
\affiliation{Kavli Institute for Particle Astrophysics and Cosmology, Stanford University, Stanford, CA 94305, USA}

\author{Ren{\'e} Doyon}
\affiliation{Institut de Recherche sur les Exoplanètes, Départment de Physique, Université de Montréal, Montréal QC H3C 3J7, Canada}

\author[0000-0002-7821-0695]{Katherine B. Follette}
\affiliation{Department of Physics and Astronomy, Amherst College, 21 Merrill Science Drive, Amherst, MA 01002, USA}

\author{Steven Goodsell}
\affiliation{Gemini Observatory, 670 N. A'ohoku Place, Hilo, HI 96720, USA}

\author[0000-0002-7162-8036]{Alexandra Z. Greenbaum}
\affiliation{Department of Astronomy, University of Michigan, Ann Arbor, MI 48109, USA}

\author[0000-0003-3726-5494]{Pascale Hibon}
\affiliation{European Southern Observatory, Alonso de Cordova 3107, Vitacura, Santiago, Chile}

\author{Patrick Ingraham}
\affiliation{Large Synoptic Survey Telescope, 950 N Cherry Av, Tucson AZ 85719, USA}

\author[0000-0002-9936-6285]{Quinn Konopacky}
\affiliation{Center for Astrophysics and Space Sciences, University of California, San Diego, 9500 Gilman Drive, La Jolla, CA 92093, USA}

\author[0000-0001-7687-3965]{James E. Larkin}
\affiliation{Department of Physics \& Astronomy, 430 Portola Plaza, University of California, Los Angeles, CA 90095, USA}

\author{Jerome Maire}
\affiliation{Center for Astrophysics and Space Sciences, University of California, San Diego, 9500 Gilman Drive, La Jolla, CA 92093, USA}

\author[0000-0002-5251-2943]{Mark S. Marley}
\affiliation{NASA Ames Research Center, MS 245-3, Moffett Field, CA, 94035, USA}

\author[0000-0002-4164-4182]{Christian Marois}
\affiliation{National Research Council of Canada Herzberg, 5071 West Saanich Road, Victoria, BC V9E
2E7, Canada}

\author{Elisabeth Matthews}
\affiliation{Department of Physics, and Kavli Institute for Astrophysics and Space Research, Massachusetts Institute of Technology, Cambridge,
MA, USA}

\author[0000-0003-3050-8203]{Stanimir Metchev}
\affiliation{Department of Physics and Astronomy, The University of Western Ontario, London, ON, N6A 3K7, Canada}

\author[0000-0001-6975-9056]{Eric L. Nielsen}
\affiliation{Kavli Institute for Particle Astrophysics and Cosmology, Stanford University, Stanford, CA 94305, USA}

\author[0000-0001-7130-7681]{Rebecca Oppenheimer}
\affiliation{American Museum of Natural History, New York, NY 10024, USA}

\author{David Palmer}
\affiliation{Lawrence Livermore National Laboratory, 7000 East Ave., Livermore, CA 94550}

\author{Lisa A. Poyneer}
\affiliation{Lawrence Livermore National Laboratory, 7000 East Ave., Livermore, CA 94550}

\author{Laurent Pueyo}
\affiliation{Space Telescope Science Institute, 3700 San Martin Drive, Baltimore MD 21218 USA}

\author[0000-0002-9246-5467]{Abhijith Rajan}
\affiliation{Space Telescope Science Institute, 3700 San Martin Drive, Baltimore MD 21218 USA}

\author[0000-0003-0029-0258]{Julien Rameau}
\affiliation{Institut de Recherche sur les Exoplan{\`e}tes, D{\'e}partement de Physique, Universit{\'e} de Montr{\'e}al, Montr{\'e}al, QC, H3C 3J7, Canada}

\author[0000-0002-9667-2244]{Fredrik T. Rantakyr\"o}
\affiliation{Gemini Observatory, Casilla 603, La Serena, Chile}

\author[0000-0003-1698-9696]{Bin Ren}
\affiliation{Department of Physics and Astronomy, Johns Hopkins University, Baltimore, MD 21218, USA}

\author[0000-0002-8711-7206]{Dmitry Savransky}
\affiliation{Sibley School of Mechanical and Aerospace Engineering, Cornell University, Ithaca, NY 14853, USA}

\author{Adam Schneider}
\affiliation{School of Earth and Space Exploration, Arizona State University, Tempe, AZ 85281, USA}

\author[0000-0003-1251-4124]{Anand Sivaramakrishnan}
\affiliation{Space Telescope Science Institute, 3700 San Martin Drive, Baltimore MD 21218 USA}

\author[0000-0002-5815-7372]{Inseok Song}
\affiliation{Physics and Astronomy, University of Georgia, 240 Physics, Athens, GA 30602, USA}

\author[0000-0003-2753-2819]{R{\'e}mi Soummer}
\affiliation{Space Telescope Science Institute, 3700 San Martin Drive, Baltimore MD 21218 USA}

\author{Melisa Tallis}
\affiliation{Kavli Institute for Particle Astrophysics and Cosmology, Stanford University, Stanford, CA 94305, USA}

\author[0000-0002-9121-3436]{Sandrine Thomas}
\affiliation{Large Synoptic Survey Telescope, 950N Cherry Av, Tucson AZ 85719, USA}

\author[0000-0001-5299-6899]{J. Kent Wallace}
\affiliation{Jet Propulsion Laboratory, California Institute of Technology, 4800 Oak Grove Dr., Pasadena CA 91109, USA}

\author[0000-0002-4479-8291]{Kimberly Ward-Duong}
\affiliation{Department of Physics and Astronomy, Amherst College, 21 Merrill Science Drive, Amherst, MA 01002, USA}

\author[0000-0003-4483-5037]{Sloane J. Wiktorowicz}
\affiliation{Remote Sensing Department, The Aerospace Corporation, 2310 E. El Segundo Blvd. M2-266, El Segundo, CA 90245 USA}

\author{Ben Zuckerman}
\affiliation{Department of Physics \& Astronomy, 430 Portola Plaza, University of California, Los Angeles, CA 90095, USA}

\begin{abstract}
    We present the first spatially resolved scattered-light images of four debris disks around members of the Scorpius-Centaurus (Sco-Cen) OB Association with high-contrast imaging and polarimetry using the Gemini Planet Imager (GPI). All four disks are resolved for the first time in polarized light and one disk is also detected in total intensity. The three disks imaged around HD 111161, HD 143675, and HD 145560 are symmetric in both morphology and brightness distribution. The three systems span a range of inclinations and radial extents. The disk imaged around HD 98363 shows indications of asymmetries in morphology and brightness distribution, with some structural similarities to the HD 106906 planet-disk system. Uniquely, HD 98363 has a wide co-moving stellar companion Wray 15-788 with a recently resolved disk with very different morphological properties. HD 98363 A/B is the first binary debris disk system with two spatially resolved disks. All four targets have been observed with ALMA, and their continuum fluxes range from one non-detection to one of the brightest disks in the region. With the new results, a total of 15 A/F-stars in Sco-Cen have resolved scattered light debris disks, and approximately half of these systems exhibit some form of asymmetry. Combining the GPI disk structure results with information from the literature on millimeter fluxes and imaged planets reveals a diversity of disk properties in this young population. Overall, the four newly resolved disks contribute to the census of disk structures measured around A/F-stars at this important stage in the development of planetary systems.
\end{abstract}


\section{Introduction}
Circumstellar debris disks around young stars are dusty remnants of protoplanetary disks \citep{wyatt2008,zuckerman2001,hughes2018}. The first evidence of a circumstellar debris disk was identified around Vega, after the Infrared Astronomical Satellite (\textit{IRAS}) observed an excess of far-IR flux, much higher than what was expected from the stellar photosphere of Vega \citep{aumann1984}. Spatially resolved imaging subsequently confirmed that infrared excesses are related to circumstellar debris disks (e.g. \citealt{smith1984,holland1998}). Infrared excess, therefore, has been a key indicator of the presence of a debris disk, and has been observed to increase with age starting at 5 Myr, peaking between 10 and 15 Myr, and then decline with age \citep{wyatt2008}. Early studies have shown that debris disks are quite common around young A stars in particular \citep{rieke2005,su2006}. Observable debris disks must continually replenish dust grains because the grains can either be accreted onto their host star or ejected from their system in relatively short timescales. Examples of processes that could sustain the dust content in debris disks include the collisional grinding of planetesimals \citep{backman1993} or a catastrophic collision of planets \citep{cameron1997}.

Scorpius-Centaurus (Sco-Cen) is the nearest ($\sim$110 -- 140 pc) OB association \citep{blaauw1946,preibisch2008,dezeeuw1999} with ages from 10--16 Myr \citep{pecaut2016,pecaut2012}, and it has proven to be a rich experimental laboratory for investigating star and planet formation. The estimated age is ideal for debris disks, as the age of the association is approximately the age when fractional infrared excess is at its highest \citep{wyatt2008}. The association has been surveyed extensively at wavelengths from optical to far-IR, enabling the identification of infrared excess sources from a uniform data set and analysis \citep{chen2014}. The targets for this study are all members of either the Lower Centarurus Crux (LCC) or Upper Centaurus Lupus (UCL) region of Sco-Cen with ages of 11--16 Myr \citep{pecaut2016}, corresponding to later parts of the era of giant and terrestrial planet formation, during which planet-disk interactions may sculpt disk structures amid planetary orbits. An example of a Sco-Cen system with both an imaged planet and resolved disk from GPI data is HD 106906 \citep{kalas2015,bailey2014}. Other examples of resolved Sco-Cen disks include the transitional disk HD 100546 \citep{augereau2001,follette2017,rameau2017} and the debris disk HD 111520 \citep{draper2016}. HD 100546 shows complex spiral arms and an inner clearing, consistent with dynamical models of planet-induced spiral structure \citep{follette2017}, while the HD 111520 disk possesses the most extreme example of an asymmetric edge-on debris disk, apparently due to 2:1 azimuthal dust density variation within the disk \citep{draper2016}.

Obtaining new images of previously unresolved disks to map disk morphologies is the main science goal of this program. Dust belts, cleared gaps, offsets, and asymmetries can be clearly measured, allowing inferences on disk dynamics and evolution (e.g. \citealt{lee2016}). Structures such as asymmetries, gaps, and clumps can encode the effects of gravitational interactions between planets and disks (e.g. \citealt{liou1999,kuchner2003,wyatt2006,quillen2006}), including planets below the current detection threshold of direct imaging. 

Scattered-light observations provide high angular resolution images for debris disks in the near-IR, similar to long-baseline near- and mid-IR interferometry mapping of disks \citep{defrere2011,absil2013}. Therefore, high contrast AO imaging is an important probe of disk structure and grain properties. In scattered light observations, disk observations are difficult because of the amount of contrast needed between the disk and bright star. Instrumental point-spread functions introduce further complications, as they spread starlight to angular separations where disks are found \citep{millar-blanchaer2016a}.

In this paper, we present Gemini Planet Imager (GPI) observations of four debris disks in Sco-Cen, all of which are imaged in scattered light for the first time. In \S \ref{targets}-\ref{datareduction}, we describe the target properties, observations, and image-processing. In \S \ref{results}, we show the images and empirical surface brightness profiles for all four disks in polarized light, along with the total intensity results when detected. In \S \ref{Discussion}, we describe the unusual properties of the HD 98363 disk, compare the new images with models in order to understand system architectures, place the four new disks in the broader context of the Sco-Cen disk population, and compare the scattered light results with ALMA millimeter dust and gas maps. In \S \ref{conclusion} we give the summary and implications of the findings.

\section{Target Properties} \label{targets}
The four targets—--HD 98363, HD 111161, HD 143675, and HD 145560–--satisfy a set of astrophysical criteria related to stellar age, spectral type, formation region, and circumstellar environment. All targets are early-type A/F stars that are members of either the Lower Centaurus Crux (LCC) or Upper Centaurus Lupus (UCL) sub-region of the Sco-Cen OB Association. Given the ${\gtrsim}$100 pc distances to Sco-Cen members based on GAIA parallaxes \citep{gaia2018}, only early type stars provide sufficient flux for the GPI wavefront sensor (R$<$9 mag). Spectral energy distributions (SEDs) provide indirect evidence of debris disks around each star based on excess emission above the level expected for the stellar photosphere. The infrared excess, $L_{IR}/L_{*}$, for three targets -- HD 143675, HD 145560, and HD 98363 -- is taken from the \cite{chen2014} study that included wavelength coverage extending from the optical to far-IR range. For the final target HD 111161, the value of $L_{IR}/L_{*}$ is taken from the \cite{mcdonald2012} study that fit SEDs covering optical to mid-IR wavelengths. Although an \textit{IRAS} 60~$\micron$ flux is measured within ${\sim}15\arcsec$ of HD 111161 (within the \textit{IRAS} beam size at this wavelength), there is a co-moving companion HIP 62488 with a separation of 13$\farcs$4 \citep{andrews2017}, making it unclear if the flux is associated with the primary, secondary, or both components. Together, the infrared excesses of the four targets range from $\sim 4 \times 10^{-4}$ to $6.4 \times 10^{-3}$ \citep{chen2014,mcdonald2012}, which are among the higher $\sim$25\% of the $L_{IR}/L_{*}$ values for UCL/LCC early-type members with Spitzer-detected excesses, but not the most extreme examples of IR excess sources (e.g. \citealt{chen2014}).

Three of the targets were included in a comprehensive SED analysis of Spitzer-detected debris disks, and all are best fit by a model of two dust belts at distinct temperatures \citep{chen2014}. Follow-up SED modeling \citep{jang-condell2015} suggested that the two dust belts in the UCL targets HD 143675 and HD 145560 are separated by a narrow gap consistent with dynamical clearing by a single planetary mass companion, although the predicted contrast and separation requirements to image the simulated companion are beyond the limitations of current high-contrast instruments such as GPI. The SED of HD 98363 was best fit by models including a grain population of crystalline silicates \citep{jang-condell2015}. Fundamental stellar and circumstellar disk properties inferred from SED models for all four systems are summarized in Table \ref{steltable1}.

By restricting the sample to stars with a common mass range, formation environment and detection of a spatially resolved disk, it is possible to investigate the diversity of disk structures present at an important phase of the development of planetary systems. The new results from this sample are combined with analogous results from GPI high-contrast imaging of other Sco-Cen A/F-stars to build a larger census of disk properties. The comparison of the disk properties in this study with other Sco-Cen members observed with GPI is given in Section \S \ref{Discussion}.

\begin{table*}[t]
    \centering
    \begin{tabular}{c c c c c c c c c c c N}
    \hline\hline
    Name & Subgroup & Sp. Type & $L_{IR}/L_*$ & $R_*$ [$R_{\odot}$] & $T_{eff}$ [K] & $M_*$ [$M_{\odot}$]& D [pc] & $M_H$ & Binary? & References\\
    \hline
    HD 98363 & LCC & A2V & 6.4 $\times$ $10^{-4}$ & 1.6 & 8830 & 1.92 & 138.6 & 1.78 & 49$\farcs$7 & 4, 5, 6, 11 \\
    HD 111161 & LCC & A3III/IV & 5.5 $\times$ $10^{-4}$ & 1.6 & 8073 & 2.4 & 109.4 & 2.05 & 13$\farcs$4 & 1, 2, 4, 5, 11, 16, 17 \\
    HD 143675 & UCL & A5IV/V & 4.1 $\times$ $10^{-4}$ & 1.3 & 8200 & 2.0 & 113.4 & 2.38 & N & 3, 4, 5, 7, 12, 14, 15 \\
    HD 145560 & UCL & F5V & 1.4 $\times$ $10^{-3}$ & 1.5 & 6500 & 1.4 & 120.4 & 2.45 & N & 1, 3, 4, 5, 13, 14 \\
    
    \hline
    \end{tabular}
    \caption{Stellar properties for the four targets within this sample. The radius of HD 143675 was found in \cite{ballering2014}, all other stellar radii were estimated with \cite{siess2000} using measured photometry and distances. \textbf{References:} 1. \cite{lieman-sifry2016}, 2. \cite{rizzuto2012}, 3. \cite{chen2014} 4. \cite{moor2017} 5. \cite{gaia2018} 6. \cite{bohn2019}, 7. \cite{ballering2014}, 8. \cite{siess2000}, 9. \cite{hog2000}, 10. \cite{pecaut2013}, 11. \cite{houk1975}, 12. \cite{houk1978}, 13. \cite{houk1982}, 14. \cite{chen2012}, 15. \cite{mittal2015}, 16. \cite{mcdonald2012}, 17. \cite{andrews2017}}
    \label{steltable1}
\end{table*}

\section{Observations} \label{observations}
\subsection{GPI Observation Modes} \label{obsmodes}
The Gemini Planet Imager (GPI; \citealt{macintosh2014}) has two main modes of operation to achieve high-contrast images: (1) a polarimetry mode consisting of a rotatable half-wave plate (HWP) and a Wollaston prism analyzer, and (2) a spectroscopy mode employing a prism and integral field unit. It is designed specifically for spatially-resolved, high-contrast observations of debris disks in the infrared (\citealt{perrin2010,perrin2015}, see Table \ref{steltable1}.)  In combination with a coronagraph, differential polarimetry efficiently rejects stellar PSF halo speckles to achieve contrasts close to the fundamental photon noise limit for polarized light from disks \citep{millar-blanchaer2016b}. Since the starlight is unpolarized, it will cancel in the different Stokes modes and enhance the detectability of the disk scattered light that is polarized. 

High contrast in spectroscopy mode is achieved through a combination of angular differential imaging (ADI) which utilizes the field rotation to disentangle stellar speckles from the disk or companion \citep{marois2006} and/or spectral differential imaging (SDI) \citep{lafreniere2007} which relies on the radial shift of speckles from the rescaling of speckle patterns as a function of wavelength compared to a fixed position for astrophysical emission \citep{marois2004}. Since SDI is most effective for objects with distinct spectral features and compact emission, disk detections presented here are solely based on ADI rather than SDI for GPI spectroscopy mode.

\begin{table*}[t]
    \begin{center}
    \begin{tabular}{c c c c c c c c c}
    \hline\hline
    Name & Obs. Mode & Date (UT) & $N$ & $t_{exp}$ (s) & Median Airmass & Seeing & $\Delta \theta$ & AO Wavefront Error [nm]\\
    \hline
    HD 98363 & Pol & 2019 February 20 & 36 & 90 & 1.204 & --- & $28.^{\circ}6$ & 137 \\
    HD 111161 & Spec & 2018 February 04 & 28 & 90 & 1.253 & $0\farcs6$ & $16.^{\circ}9$ & 115 \\
     & Pol & 2018 March 10 & 76 & 60 & 1.288 & $0\farcs8$ & $38.^{\circ}0$ & 144 \\
    HD 143675 & Spec & 2018 April 08 & 53 & 60 & 1.008 & 0$\farcs$7-1$\arcsec$ & $94.^{\circ}3$ & 147\\
     & Pol & 2018 April 08 & 16 & 60 & 1.010 & 0$\farcs$7-1$\arcsec$ & $21.^{\circ}0$ & 170\\
    HD 145560 & Spec & 2018 August 12 & 38 & 60 & 1.040 & $0\farcs5$ & $36.^{\circ}1$ & 150 \\
     & Pol & 2018 August 12 & 28 & 60 & 1.068 & $0\farcs5$ & $17.^{\circ}6$ & 153 \\
    
    \hline
    \end{tabular}
    \caption{Summary of observations, where field rotation applies to spectral exposures only. $N$ refers to the number of exposures.}
    \label{obstable}
    \end{center}
\end{table*}

\subsection{GPI Observations} \label{targetobs}
The observations were obtained through two programs that had distinct data acquisition strategies, although both programs utilized the $H$-band filter which provides a balance between AO performance and thermal sky background. Three targets---HD 111161, HD 143675, and HD 145560---were observed as part of the Gemini Planet Imager Exoplanet Survey (GPIES) project (GS-2014B-Q-500) which included a disk survey component \citep{esposito2019}. All the observations preseneted here except for HD 98363 include a spectral sequence of 38---53 exposures of 59.65 s or 88.97 s each. Sequence lengths were adjusted due to conditions. The total number of spectral mode exposures and cumulative field rotation ($\Delta \theta$) obtained over these exposures is recorded for each target in Table \ref{obstable}, along with environmental conditions of average seeing (from 0$\farcs$5 to variable), the wavefront error determined by the spot offset measurements recorded by the Shack-Hartmann wavefront sensor, and the airmass at the midpoint of the sequence.

In addition to the spectral data, the targets in the GPIES disk campaign were observed in polarimetry mode by obtaining a sequence of 60~s images during which the HWP cycled through rotation angles of 0.0, 22.5, 45, and 67.5 degrees. In most cases, the polarimetry mode data were taken immediately following the spectral mode data; however, one target (HD 111161) required two separate nights to acquire both modes of observations. Table \ref{obstable} lists the number of individual polarization images taken for each target and the exposure times utilized in this mode. Since field rotation is most important for spectral images where KLIP-ADI is applied, it is not listed for polarized intensity observation sequences.

The final target, HD 98363, was observed in a follow-up program to the GPIES disk campaign (GS-2019A-Q-109) that focused on disk detection and employed only the polarimetry mode. Given the somewhat fainter stellar magnitude for this target with the farthest distance, the individual exposure times were set to 90~s to increase the signal-to-noise ratio (S/N), while remaining sufficiently short to prevent smearing of the PSF during exposures taken near transit when the field rotation rate is highest. The observation date, environmental conditions, and number of exposures for HD 98363 are listed in Table \ref{obstable} along with the three targets observed in GPIES.

\section{Image Processing} \label{datareduction}
The data were reduced with the GPI Data Reduction Pipeline (see \citealt{perrin2014, perrin2016, wang2018} for details). The raw data were dark subtracted, cleaned of correlated noise, and corrected for bad pixels. Spectral data were subsequently flexure-corrected and wavelength-calibrated with an Ar lamp exposure taken before the science observation sequence. After initial processing, the polarimetry data were flexure-corrected and combined into a polarization datacube. Each polarization datacube was divided by a polarized flat field and corrected for non-common path errors through a double differencing algorithm \citep{perrin2015}. The instrumental polarization was estimated by the stellar polarization in each datacube. This was done by measuring the mean normalized difference of pixels with separations that varied with each dataset. For all datasets, the full range of separations were between 8 and 17 pixels from the location of the star in the image \citep{wang2014}. Instrumental polarization was then subtracted from each pixel, scaled by the pixel total intensity \citep{millar-blanchaer2015}. The region used to measure the instrumental polarization was located just outside the edge of the focal plane mask of the coronagraph, where instrumental polarization is expected to be at a maximum.

The polarimetric and spectral datacubes were also corrected for geometric distortion, smoothed with a Gaussian kernel ($\sigma$ = 1 pixel) and combined into a Stokes datacube as demonstrated in \cite{perrin2015}. The Stokes datacube was then converted to a radial Stokes cube \citep{schmid2006}. Here, a positive $Q_{\phi}$ corresponds to polarized intensity vectors oriented perpendicular to a line connecting the star to an individual pixel while negative values correspond to parallel vectors. Single-scattering debris disks are not expected to produce polarized intensity vectors oriented $\pm 45^{\circ}$ to the same line, suggesting that a $U_{\phi}$ image should not have any disk flux and will only have noise. Using the flux of the four satellite spots in each image, flux calibration for the polarimetric and spectral datacubes was performed as discussed in \cite{hung2015}.

Both the polarization and spectral datacubes were also processed separately using the pyKLIP \citep{wang2015} implementation of the Karhunen-Lo{\`e}ve Image Projection (KLIP) algorithm \citep{soummer2012} with Angular Differential Imaging (ADI, e.g. \citealt{marois2006,lafreniere2007}) in order to search for the disks in total intensity light. For PSF subtraction with pyKLIP-ADI, 5 Karhunen-Lo{\`e}ve modes were used. To determine the sensitivity to point source companions, contrast curves are generated for the spectral observational datasets of HD 111161, HD 143675, and HD 145560. For HD 98363, a contrast curve was generated for the polarimetry dataset. Per \cite{wang2015}, assuming azimuthally symmetric noise, pyKLIP calculates the 5$\sigma$ noise level at a range of radial separations throughout the image. To assess sensitivity to planets, 12 fake planets at known brightness are injected into the pyKLIP-reduced images. The brightness of the planets scales to the detection threshold at different radial separations. The images are passed through pyKLIP once again and the flux of each injected planet is retrieved to calculate the final calibrated contrast curves. All contrast curves were calculated using a pyKLIP reduction using 30 KL modes.

\section{Results} \label{results}
\subsection{Disk Images in Polarized Light and Total Intensity}
\begin{figure*}[t]
\centering
\includegraphics[width=6.5in]{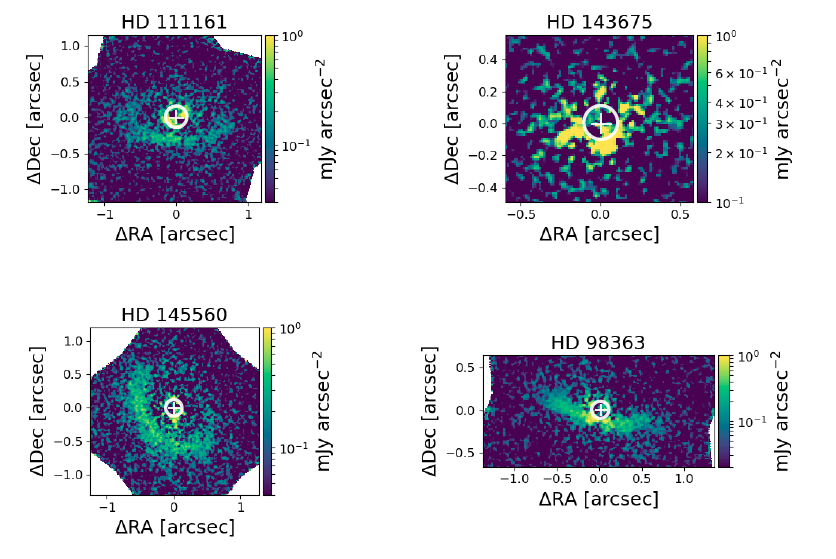}
\caption{Stokes $Q_{\phi}$ images, with the star located at coordinate (0,0). Images are in units of mJy arcsec$^{-2}$ and presented on a log scale stretch.}
\label{polimages}
\end{figure*}

The polarized intensity $Q_{\phi}$ image for each target is shown in Figure \ref{polimages}, revealing spatially resolved structures for each debris disk. HD 143675 has detected disk flux restricted to within ${\sim}0\farcs4$ from the host star, and HD 98363 is the most extended disk, with the diffuse eastern side detectable to ${\sim}0\farcs9$ from the star. The HD 111161 and HD 145560 disks show ring-shaped structures that are less inclined and more diffuse than the nearly edge-on systems. HD 111161 is moderately inclined, with the south edge being the front edge assuming forward-scattering grains. The image presents a ring with a dust-depleted inner region. HD 145560 presents the most face-on geometry and radially broad structure, with portions of the back side of the disk visible but significantly fainter than the front, southwest edge. The image of HD 145560 shows a surface brightness deficit directly north of the star, but given the generally low surface brightness and poor S/N, it is unlikely that this is a real dust gap.

\begin{figure}[t]
    \centering
    \includegraphics[width=\columnwidth]{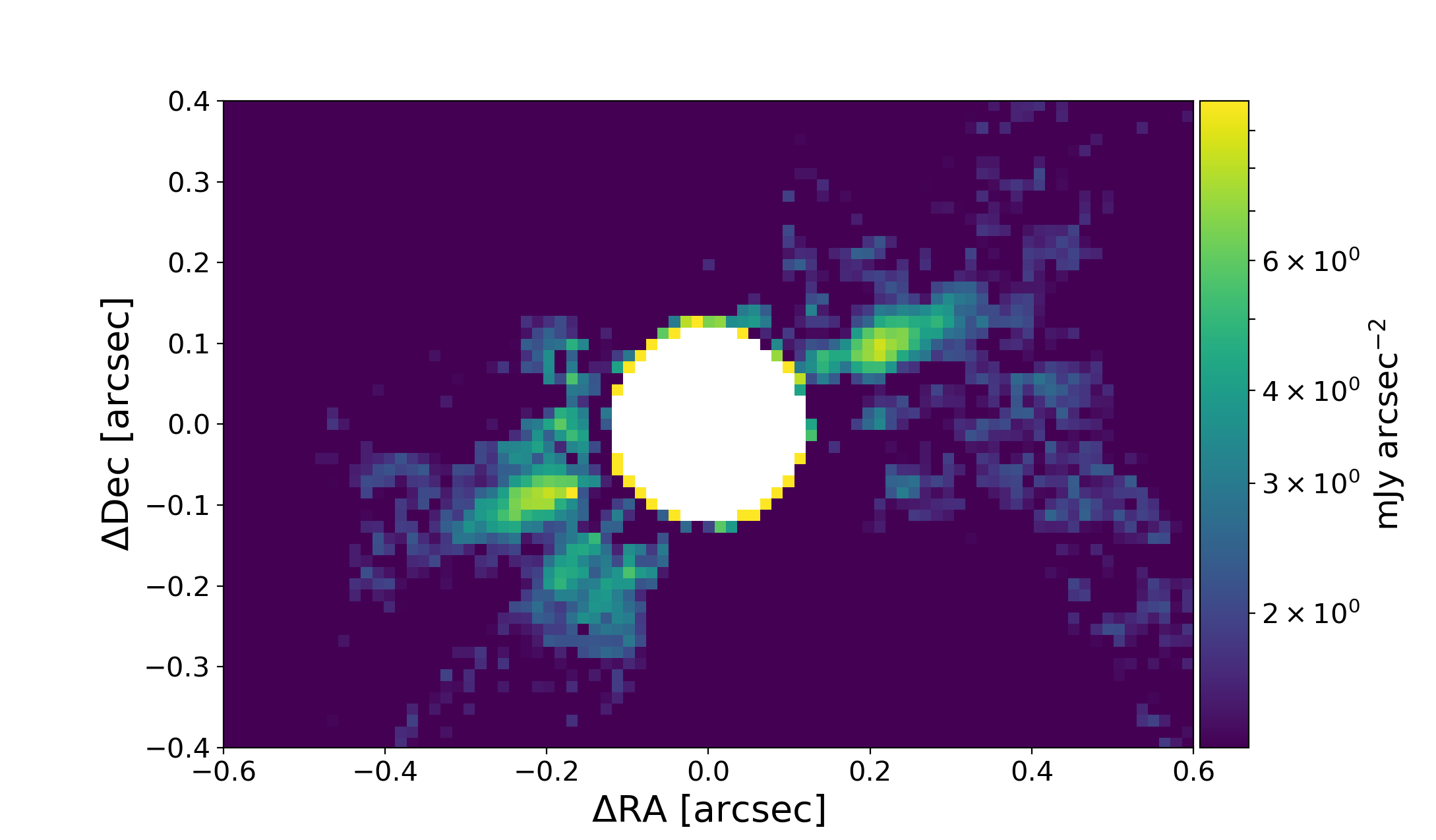}
    \caption{Total intensity image for HD 143675, made using pyKLIP-ADI with 5 KL modes. The central white region represents the extent of the focal plane mask of the coronagraph. Image is in units of mJy arcsec$^{-2}$ and presented on a log scale stretch.}
    \label{143675totint}
\end{figure}

Of the four targets, only the HD 143675 disk is detected with statistical significance in total intensity, as shown in Figure \ref{143675totint}. From the total intensity image, the HD 143675 disk has an edge-on geometry and is the most compact. The nearly edge-on geometry of HD 143675 is the most favorable case for detection with ADI, unlike a diffuse or face-on structure. As summarized in Table \ref{obstable}, HD 143675 also had the largest range of field rotation and largest number of spectral mode exposures (but not the most total integration time), enhancing the capacity to resolve this disk relative to the other targets in total intensity. Although HD 98363 also has a similar edge-on geometry, the observation did not have sufficiently high enough field rotation for a detection after pyKLIP-ADI was applied.

\subsection{Disk Morphologies and Surface Brightness Profiles}
By measuring the surface brightness along the disk spine (the midpoint of the vertical brightness distribution along the disk), the brightness and morphological asymmetry of a debris disk can be assessed. The disk geometry and morphology affect the measurement and interpretation of a surface brightness profile. The inclination places limits on the observable scattering angles, with edge-on systems generally restricted to a narrow range of scattering angles less than 90$^{\circ}$ on either side of the star. Scattering angles are estimated using an estimated $R_{\mathrm{in}}$ calculated from the known distance to the star and the observed spatial extent of the disk in the images. The association of a given disk image position with a scattering angle is predicated on the assumption of a symmetric disk centered on the star, making asymmetric disks more difficult to model. 

For each disk, we characterized its symmetry and surface brightness from the GPI data. To determine the brightness of the disk, we rotate each disk image to be approximately horizontal to measure its surface brightness profile. With rectangular apertures 7 pixels wide in the vertical direction \citep[see][]{draper2016} and 5 pixels in the horizontal direction centered on the disk midplane, we measure the surface brightness profile assuming the debris disk is a circular ring centered around the host star. The signal within each aperture is summed. To determine the uncertainties, apertures of the same size are placed in the same region where the disk is located but in the $U_{\phi}$ polarization image. Assuming forward-scattering Mie grains, the $U_{\phi}$ polarization state is not expected to contain any disk flux. The signal within these apertures is summed, and a common error is found by finding the standard deviation of the background apertures.

\begin{figure*}
    \centering
    \includegraphics[width=7in]{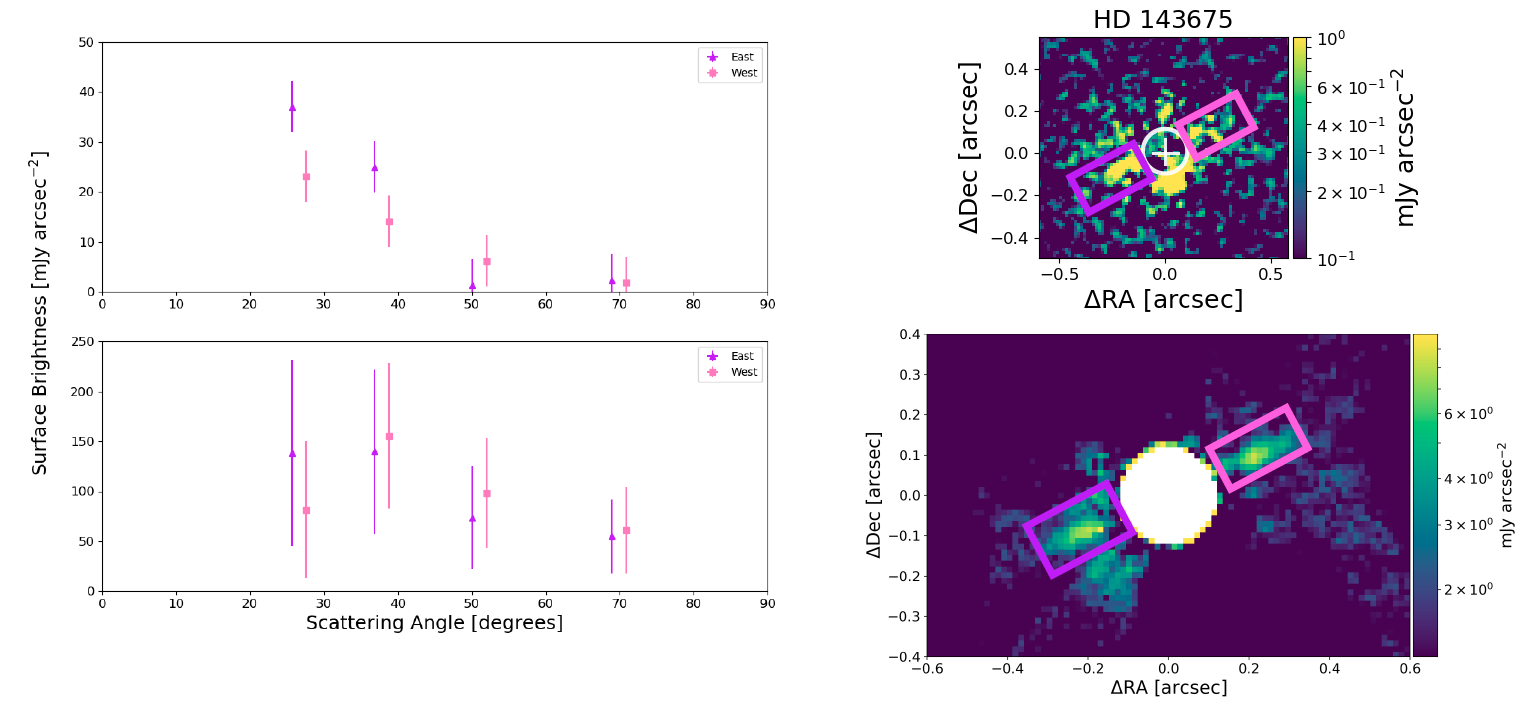}
    \caption{Surface brightness profiles measured from HD 143675 in both total and polarized intensity. The colored boxes overlaid on the image mark the regions in which measurement apertures were placed. In the total intensity surface brightness profile, the SE and NW sides appear to be fairly symmetric within uncertainties. The white region represents the extent of the focal plane mask of the coronagraph. For the polarized intensity surface brightness profile, the SE side appears marginally brighter, but this result should be treated with caution due to the close proximity of the disk to the focal plane mask.}
    \label{hd143675SB}
\end{figure*}

For the nearly edge-on and compact HD 143675, the disk reaches projected separations of $\sim$0$\farcs$35 in polarized intensity and $\sim$0$\farcs$45 in total intensity. With a distance of 139.20 pc, these projected separations are equivalent to a range of 49-63 AU which is typical for an outer radius of a debris disk \citep{esposito2019}. The largest observable scattering angles are approximately $-70^{\circ}$ and +$70^{\circ}$ from the star based upon the width of the apertures and the edge-on geometry of the system. These scattering angles are located close to the ansae of the ring (assuming we detect the true ansae at the outer extent of the disk emission). Meanwhile, angles less than $25^{\circ}$ are blocked by the focal plane mask. The surface brightness profiles for the SE and NW sides of the HD 143675 disk in polarized and total intensities are shown in Figure \ref{hd143675SB}. Within the uncertainties, the phase functions are symmetric in both polarized and unpolarized light. Self-subtraction from ADI is not expected to significantly introduce asymmetric features into an intrinsically symmetric disk image. In addition, a conservative number of Karhunen-Lo{\`e}ve modes were selected (KL = 5) to support higher throughput \citep{soummer2012}.

\begin{figure*}
    \centering
    \includegraphics[width=7in]{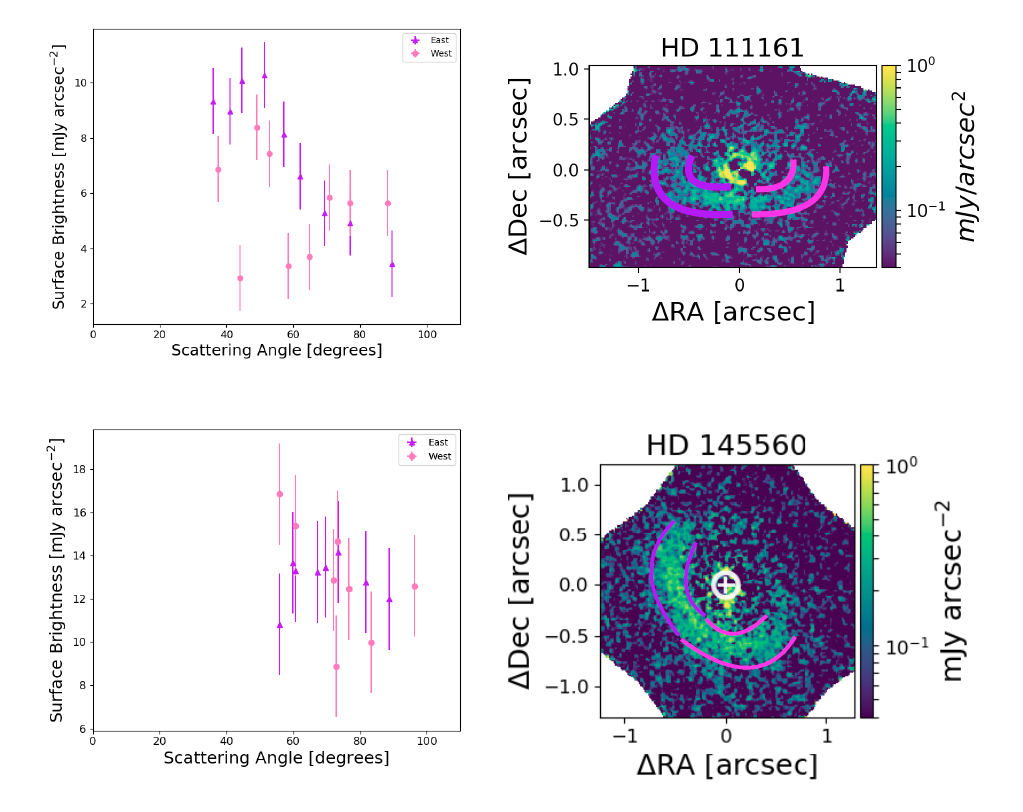}
    \caption{Scattering phase functions measured from HD 111161 (top) and HD 145560 (bottom) in polarized intensity. The colored boxes overlaid on the image mark the regions in which measurement apertures were placed. For HD 111161, a couple measurements suggest tentative asymmetric structure, but caution must be taken due to the low intrinsic brightness of the disk. For HD 145560, the E and W sides appear to be fairly symmetric.}
    \label{hdinclinedSB}
\end{figure*}

For the HD 145560 disk, the inclination enables partial access to the back side of the disk, but primarily the front NE and SE edges were measured and plotted in Figure \ref{hdinclinedSB}; the NW and SW back edges (angles $\gtrsim98\degr$) had S/N $\lesssim3$, which is less than the S/N of one measurement at ${\sim}97\degr$ along the SW edge, where the back edge of the disk begins. Because of the low S/N, surface brightness is not measured for the back edge of the disk.

For HD 111161, the front edge of the disk were measured and the results are shown in Figure \ref{hdinclinedSB}. Similar to the edge-on HD 143675 disk, the HD 145560 and HD 111161 disks appear to have symmetric surface brightness profiles within the capacity to measure differences in these discovery images. 

\begin{figure*}
    \centering
    \includegraphics[width=0.8\textwidth]{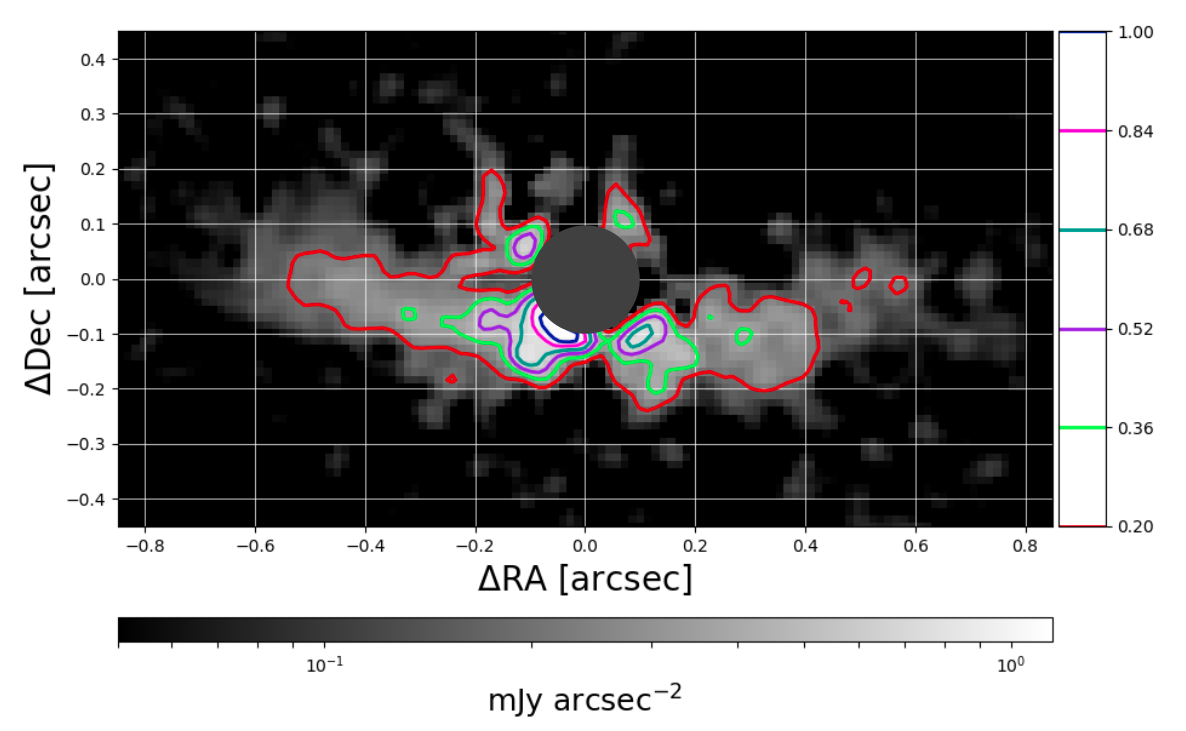}
    \caption{Surface brightness contours overlaid onto the horizontally rotated image of HD 98363 with the star at coordinate (0,0). The NE side appears to be brighter over a larger region than the SW side. Additionally, the NE side appears more radially extended than the SW side. The gray circle indicates the location of the focal plane mask.}
    \label{hd98363contour}
\end{figure*}

In contrast, the HD 98363 disk exhibits a different morphology from the other three disks, with an indication of an asymmetric structure and brightness distribution, as shown in Figure \ref{hd98363contour}. Surface brightness contours from 0.2 to 1.0 mJy arcsec$^{-2}$ were overlayed on the image of the disk to highlight the radially more extended and brighter northeast side. Due to the asymmetric shape, determining a unique scattering angle per position is more complex: unlike the other three disks, the assumption of a circular ring is not valid for HD 98363. In addition, the variable projected extent of the disk makes the definition of a consistent aperture for a brightness measurement difficult; for these reasons, a surface brightness profile is not calculated for this source and is deferred for a later study. The empirical results on the disk structures for the four targets are compared with the broader Sco-Cen disk population in \S \ref{scocenpop}.

\subsection{Contrast Curves and Planet Detection Limits}
\begin{figure*}
    \centering
    \includegraphics[width=\textwidth]{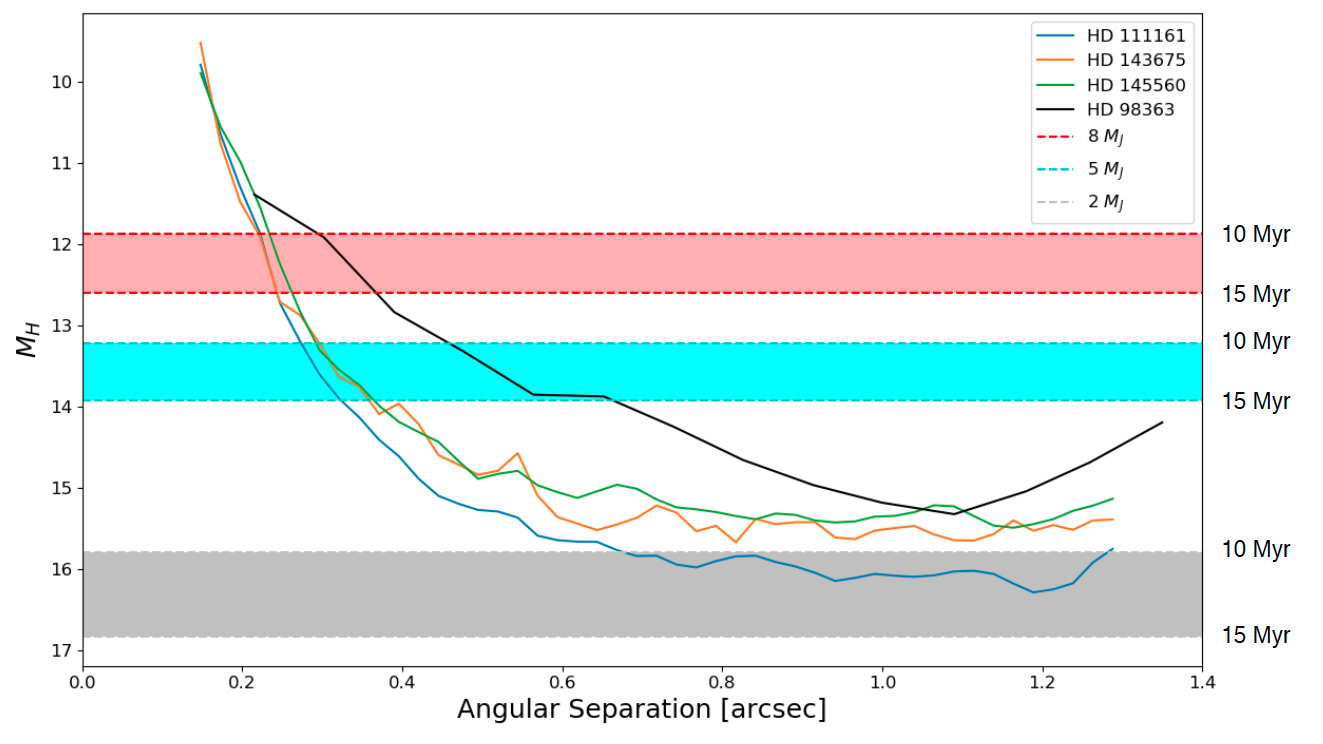}
    \caption{\textbf{Calibrated contrast curves converted to absolute \textit{H} magnitude. The absolute H-band limits are converted into planet mass limits using the COND03 \citep{baraffe2003} evolutionary models and ages of 10Myr and 15Myr. 5 $M_J$ companions would have been easily detected over most of the separation range covered in the GPI observations, while 2 $M_J$ is below the detectable limit for all but one of the targets over most of the separation range.}}
    \label{contrastcurve}
\end{figure*}

As discussed in \S \ref{datareduction}, contrast curves were generated for the polarimetry dataset of HD 98363 and the spectral datasets of HD 111161, HD 143675, and HD 145560 (Figure \ref{contrastcurve}, left). From the calibrated 5$\sigma$ contrast curves, the relative \textit{H} magnitude as a function of radial separation can be found. From the relative \textit{H} magnitudes and the reported \textit{H} magnitudes of the target stars and their distances, the upper limit absolute magnitude as a function of radial separation can be calculated and is shown in Figure \ref{contrastcurve} (right) for HD 111161, HD 143675, HD 145560, and HD 98363. By applying 10 Myr and interpolated 15 Myr COND03 evolutionary models \citep{baraffe2003}, we find that in all objects in our sample, we would not be able to detect any substellar companion with a mass of $\lesssim 2 M_{J}$.

\section{Discussion} \label{Discussion}
\subsection{The Unique HD 98363 System}
Among the debris disk systems resolved in this study, HD 98363 is an exceptional case with a $\sim$7000 AU co-moving secondary companion Wray 15-788 that also has a spatially resolved disk \citep{bohn2019}. A small set of resolved primary debris disks with imaged stellar or substellar companions are known (e.g. HD 106906, \citealt{kalas2015}), however the HD 98363 system is unique with the detection of two resolved disks. The infrared images of each component disk from GPI or SPHERE already show intriguing differences--- misaligned inclinations for the two disks, asymmetries in the HD 98363 disk, and a gap with the possibility of two belts in Wray 15-788 system \citep{bohn2019}. The asymmetric HD 98363 disk has some structural similarities to the HD 106906 system (\citealt{kalas2015} and Figure \ref{surveyfigure}) which has a wide orbit imaged planetary mass companion \citep{bailey2014}. Another unusual aspect of this double-disk system is the presence of $H\alpha$ emission \citep{wray1966,henize1976} in the secondary at a level suggesting active accretion and an earlier evolutionary state for the disk, making this a rare example of a mixed-state system, since the primary has no $H\alpha$ emission. \cite{bohn2019} estimated an $L_{IR}/L_*$ of $\gtrsim$ 0.27 for Wray 15-788, while the $L_{IR}/L_*$ is $6.4 \times 10^{-4}$ for HD 98363, further suggesting a mixed-state system of a debris disk and a transition disk.

To estimate the HD 98363 disk inclination so that it can be compared with the secondary disk, we modeled the  disk with the radiative transfer code MCFOST \citep{pinte2006}. For the modeling process, grain properties ($a_{\mathrm{min}}$, $a_{\mathrm{max}}$, $\alpha_{\mathrm{in}}$, $\alpha_{\mathrm{out}}$, and porosity) were fixed, while morphological properties ($i$, disk position angle, $R_{\mathrm{in}}$, $R_{\mathrm{out}}$, $R_{\mathrm{c}}$, and dust mass) were left as free parameters for a Markov Chain Monte Carlo (MCMC) sampling process as in \cite{esposito2019}. Only the Stokes $Q_{\phi}$ image was compared with the model equivalent. For grain properties, $a_{\mathrm{min}}$ was set to 2.00 $\mu$m, $a_{\mathrm{max}}$ was set to 1000.0 $\mu$m, $\alpha_{\mathrm{in}}$ was set to 2.0, $\alpha_{\mathrm{out}}$ was set to -3.0, and porosity was set to 1.0. Astrosillicate grains \citep{draine1984} were assumed for the grain composition, and Mie scattering theory \citep{mie1908} was used for all model scattering properties. MCMC samplings using the \texttt{emcee} Python package used 1 temperature, 126 walkers, and 1000 iterations per walker. From the MCFOST models, we conservatively constrain an inclination of $i \sim 75-85\degr$. Due to the asymmetry in the disk, we do not report values for $R_{\mathrm{in}}$, $R_{\mathrm{out}}$, and $R_{\mathrm{c}}$ as they cannot be reasonably constrained with this basic model. Compared to the $i = 21\degr \pm 6\degr$ of Wray 15-788 \citep{bohn2019}, the $i \sim 75-80\degr $ of HD 98363 is evidence of a stellar binary system with misaligned circumstellar disks.

The $\Delta i \sim 60\degr$ misalignment in inclinations for the HD 98363/Wray 15-788 system can be compared to other examples of multiple-component systems of younger protoplanetary disks in which each disk was spatially resolved. The HD 98363/Wray 15-788 binary system is very similar in misalignment to the HK Tau system \citep[see][]{koresko1998,stapelfeldt1998}. \cite{jensen2014} found that the misalignment between the two components of HK Tau was estimated at $60-68\degr$. In the case of HK Tau, the disk inclinations relative to the binary orbital plane are substantial for at least one of the circumstellar disks in the system, although the wide $\sim$7000 AU separation prevents an estimate of the HD 98363/Wray 15-788 orbital plane orientation. Also in the Taurus moving group, \cite{roccatagliata2011} found that the binary system of Haro 6-10 is also highly misaligned, with the difference in inclination estimated at $\sim 70\degr$. The T Tau triple system \citep{dyck1982} has three components with the northern component ejecting mass outflows \citep{bohm1994} suggesting a face-on circumstellar disk. \cite{duchene2005} spatially resolved a circumstellar disk around the southern component of the T Tau system, and found evidence of an edge-on disk, suggesting that the circumstellar disks around T Tau and T Tau S are even more highly misaligned than the HD98363/Wray 15-788 system \citep{skemer2008,ratzka2009}.

From numerical simulations, misalignment in disks of binary systems is expected to occur, with the most significant misaligned disks occurring with binary separations greater than 100 AU \citep{batygin2012,bate2000}. The mutual alignment or misalignment of stellar spin axes in binary pairs can also be used as a record of inclinations of binary systems. Although only a limited number of double disk systems have been spatially resolved, population studies of spin axes in binaries with separations ranging from a few AU to $\sim$1000 AU have been performed. \cite{hale1994} calculated mutual equatorial inclinations for a large sample of binary systems and found a positive trend with $\Delta i$ increasing as a function of the semimajor axis of the binary system. For the $\sim7000$ AU wide HD 98363/Wray 15-788 binary system, a $\Delta i \sim 60\degr$ is consistent with the general trend of mutual inclination versus semimajor axis in \cite{hale1994}.

The HD 98363/Wray 15-788 system represents an important type of system to explore binary-disk interactions in which each component disk is resolved. A key astrophysical question is whether or not the HD 98363 asymmetry and the Wray 15-788 double-belt structure are caused by the external dynamical perturbation of the other star in the system. Since the pair is widely separated at the current epoch, an eccentric orbit (e.g., $e >$ 0.7) would be sufficient to have an apoastron distance comparable to the disk radius which would cause a strong dynamical perturbation. Numerical simulations investigating the consequences on disk structure due to a stellar flyby perturbation during periastron suggest that an asymmetric debris disk can result from the dynamical interaction \citep{larwood2001}. The binary mass ratios explored in the simulation are similar to that of the HD 98363/Wray 15-788 system, and the dynamical model showed that a close, non-coplanar stellar encounter could rearrange the orbital elements of disk particles to generate an asymmetric structure that is vertically flat and radially extended in one direction, but radially truncated and vertically distended in the opposite direction \citep{larwood2001} --- analogous to the HD 106906 disk and with similarities to the HD 98363 disk.

\subsection{System Architectures} \label{modeling}
Our spatially resolved disk images can be compared with both scattered-light models generated to fit those images and blackbody models designed to fit SEDs in order to develop an overview of the disk architectures. Three of the targets -- HD 111161, HD 143675, and HD 145560 -- have been analyzed with both types of modeling and are considered in this section. The key parameters that characterize the disk geometries are the radii associated with the locations of dust rings.

For SED fitting, either a single temperature blackbody emission component or a pair of blackbody components with different temperatures is added to the stellar photospheric emission to match the unresolved photometry of the entire system, and then the blackbody temperatures are converted to dust belt radii $R_{1}$ and, if there is a second component, $R_{2}$. It is important to note, however, that emitting dust in debris disks is typically overheated and that dust belt radii estimated from blackbody temperatures are typically underestimated by factors of $\sim$2 and $\sim$4 for A- and F- stars respectively \citep{pawellek2015}. Despite this potential underestimation, for this comparison, we adopt the radius results from the \cite{mcdonald2012} and \cite{jang-condell2015} SED fits that are summarized in Table \ref{modeloutputs}. For HD 111161, \cite{mcdonald2012} report a dust temperature without uncertainties.

In scattered-light modeling, simulated images are generated by modeling the three-dimensional distribution of micron-sized dust grains and then computing the intensity of scattered starlight at each point in the disk (in this case using the radiative transfer modeling code MCFOST and Mie theory; \citealt{pinte2006}). The simulated images are then iteratively compared with spatially resolved maps from high-contrast imaging through MCMC sampling. This provides a quantitative estimate of the observed disk radius while taking into account geometric projection and scattering phase function effects. For the three disks considered here, we adopt the radii presented in \cite{esposito2019}. Their median-likelihood values for the inner and outer scattered-light radii, $R_{\mathrm{in}}$ and $R_{\mathrm{out}}$, are quoted in Table \ref{modeloutputs} with uncertainties corresponding to the 34\% confidence intervals of the MCMC posterior distributions. We also use the disk inclination to translate to the observed view of each of the three disks. The GPI inner working angle is determined by the radius of the focal plane mask of the coronagraph, which is listed in AU in Table \ref{modeloutputs}. The SED-based dust belt locations $R_{1}$ and $R_{2}$ are generally interior to this limit, except for the case of $R_2$ for HD 145560. For HD 98363, the MCMC modeling was performed to estimate the disk inclination for comparison with the primary disk. Due to the asymmetric nature of the HD 98363 disk, the values for $R_{\mathrm{in}}$ and $R_{\mathrm{out}}$ are systematically biased with a symmetric disk model, and therefore the values are not reported.

\begin{figure*}[t]
    \centering
    \includegraphics[width=\textwidth]{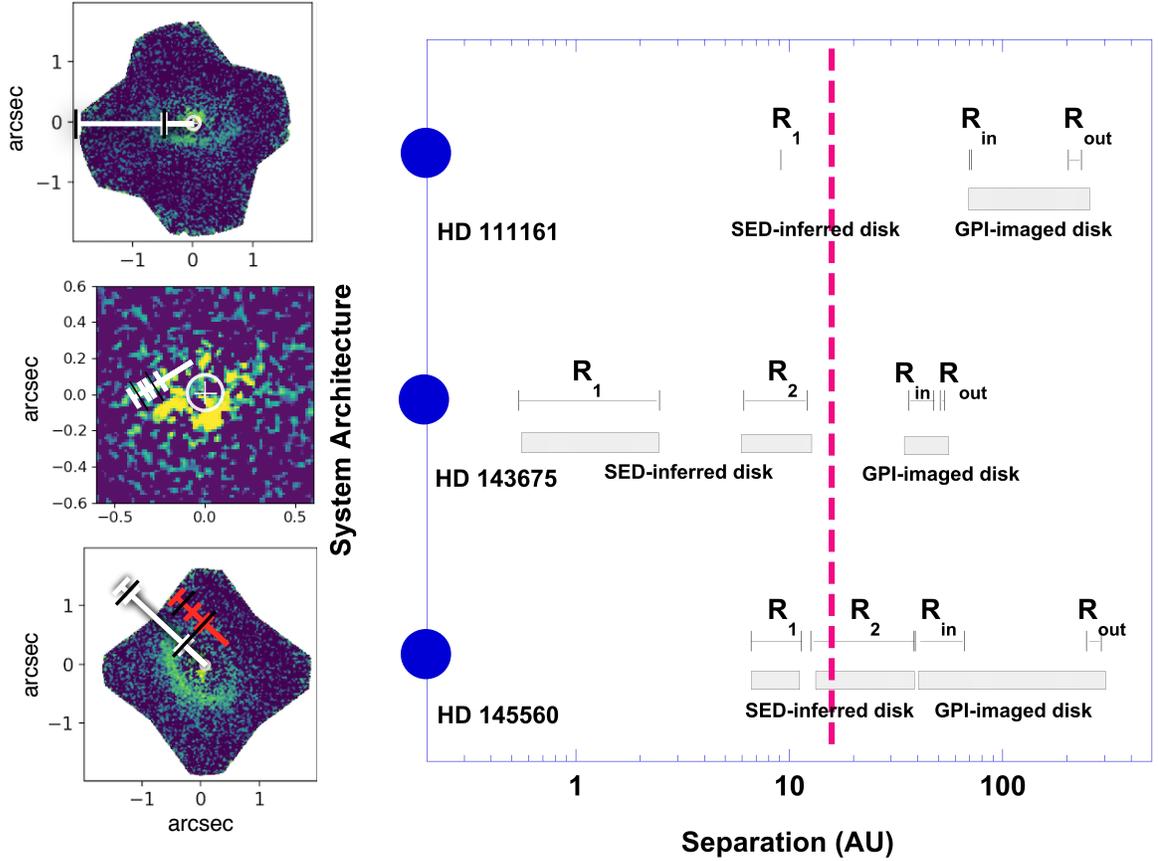}
    \caption{\textit{Left:} GPI H-band polarized light images of HD 111161 (top), HD 143675 (middle), and HD 145560 (bottom), with the radial line showing the inclination-projected locations of the radiative transfer model fit \citep{esposito2019} inner and outer radii, as explained in \S \ref{modeling}. The black lines and surrounding white bars are the quoted values and associated uncertainties respectively. For HD 145560, the black lines and surrounding red bars are the median likelihood $R_{\mathrm{in}}$, $R_{\mathrm{out}}$, and associated uncertainties respectively as reported in \cite{lieman-sifry2016}. \textit{Right:} Schematic diagrams of the same three disks that include the estimates of locations of a single dust belt \citep{mcdonald2012} or two dust belts \citep{jang-condell2015} based on SED-fitting of unresolved photometry. The \cite{mcdonald2012} study did not report uncertainties. The vertical dashed red line indicates the separation corresponding to radius of the focal plane mask; the GPI images cannot directly resolve structures interior to this limit.}
    \label{modelSED}
\end{figure*}

The ranges obtained for $R_{\mathrm{in}}$ and $R_{\mathrm{out}}$ from the scattered-light models are shown overlaid on the GPI images in the left column of Figure \ref{modelSED}. $R_\mathrm{in}$ is constrained to be well outside the GPI inner working angle in the cases of HD 111161 and HD 145560. For the HD 143675 disk, its nominal $R_{\mathrm{in}}$ is outside the GPI inner working angle but its edge-on and radially compact nature meant that a smaller $R_{\mathrm{in}}$ near 15 au could not be ruled out with $>3\sigma$ confidence \citep{esposito2019}. The $R_{\mathrm{out}}$ of HD 143675 is consistent with the edge of the detectable disk in the GPI image, while the HD 111161 and HD 145560 $R_{\mathrm{out}}$ values extend beyond the GPI field-of-view. Additional information on the outer disks from ALMA is given in \S \ref{ALMAcomp}. 

A schematic diagram showing the system architectures for HD 111161, HD 143675, and HD 145560 is given in Figure \ref{modelSED}. The values of $R_{in}$ and $R_{out}$ are shown based on the confidence intervals given in Table \ref{modeloutputs}. The temperature from \cite{mcdonald2012} was converted into a dust location $R_{1}$ based on the effective temperature and radius of the star given in Table \ref{steltable1}. The results of the two-temperature SED-fits for HD 143675 and HD 145560 for the dust belt locations $R_{1}$ and $R_{2}$ are indicated, with $\pm$3$\sigma$ uncertainties included \citep{jang-condell2015}. For HD 111161, the exact limits of the inner and outer radii from the MCFOST model fit to the data have significant uncertainties, but the GPI-imaged structure is at larger scales than the dust belt or belts inferred from SED-fitting (even if the dust belt radius from \cite{mcdonald2012} is underestimated), and the SED analysis indicates that the interior portion of the GPI-imaged disk is not entirely clear of material. For HD 143675, $R_1$ could represent a distinct dust population from $R_2$ and the radii inferred from scattered light imaging. Due to the possible underestimation of $R_2$, it is ambiguous whether or not the dust population located at $R_2$ is distinct from the dust population inferred by the GPI scattered light images. Finally, for HD 145560, the distinction between the radii inferred from SED fitting and from scattered light imaging is ambiguous, due to the possible underestimation of $R_1$ and $R_2$. In this case, it is possible that GPI is imaging the same dust population as inferred from the SED. Taken together, the photometry, images and models suggest the disks could have a range of dust populations, from potentially one in HD 145560 to as many as 3 for HD 143675.

\begin{table*}[t]
    \centering
    \begin{tabular}{c c c c c}
        \hline
        \hline
        Parameter & HD 111161 & HD 143675 & HD 145560 & Ref\\ \hline
        $R_1$ [AU] & 9.13 & 1.5$\pm$0.32 & 9.62$\pm$1 & 1, 2\\
        $R_2$ [AU] & --- & 9.12$\pm$1 & 24.9$\pm$4.5 & 2 \\
        $R_{\mathrm{in}}$ [AU] & 71.4$^{+0.5}_{-1.0}$ & 44$^{+3.5}_{-7.6}$ & 68.6$^{+2.9}_{-1.3}$ & 3\\
        $R_{\mathrm{out}}$ [AU] & 217.9$^{+15.5}_{-15.3}$& 52.1$^{+1.4}_{-1.0}$ & 224.0$^{+27.2}_{-10.8}$* & 3\\
        Inclination [deg] & 62.1$^{+0.3}_{-0.3}$ & 87.2$^{+0.6}_{-0.7}$ & 43.9$^{+1.5}_{-1.4}$ & 3\\
        IWA [AU] & 10.94 & 11.34 & 12.04 & 4\\\hline
        
    \end{tabular}
    \caption{Radius estimates from SED fitting \citep{mcdonald2012,jang-condell2015} compared to radius estimates from MCFOST modeling \citep{esposito2019}. *For HD 145560, \cite{esposito2019} presents a lower limit for $R_{\mathrm{out}}$ of 196.2 AU for a 99.7\% confidence interval. \textbf{References:} 1. \cite{mcdonald2012}, 2. \cite{jang-condell2015}, 3. \cite{esposito2019}, 4. \cite{macintosh2014}}
    \label{modeloutputs}
\end{table*}

\subsection{Compilation of Infrared Scattered Light Disk Properties in Sco-Cen} \label{scocenpop}
\begin{figure*}
    \centering
    \includegraphics[width=0.8\textwidth]{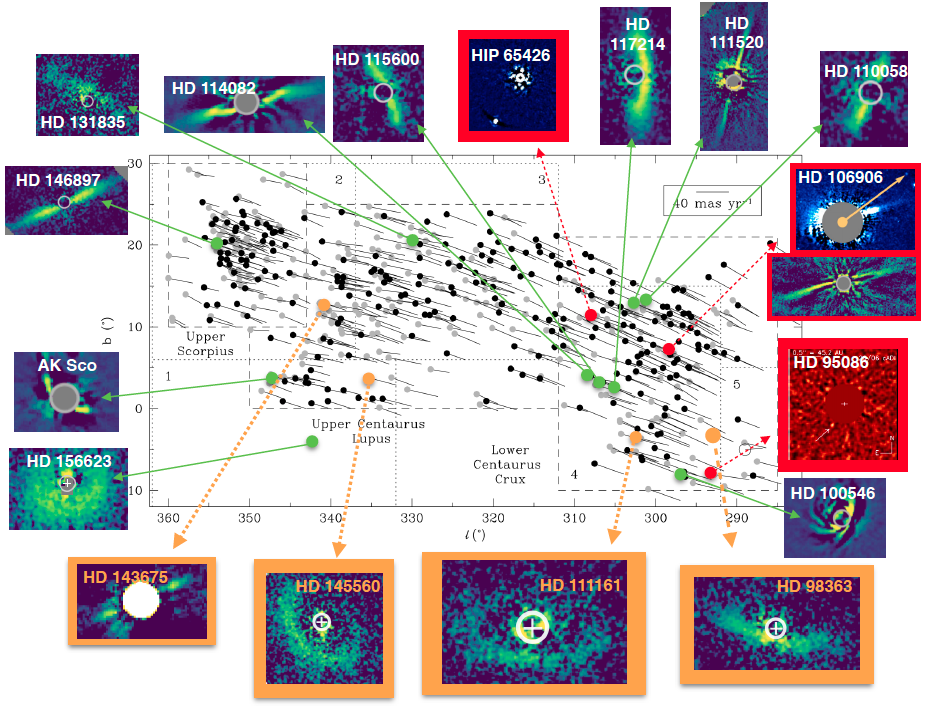}
    \caption{Gallery of resolved scattered light disks and imaged giant planets in Scorpius-Centaurus. Green points are A and F systems with resolved scattered light debris disks. Red points are A and F systems with imaged giant planets, and in the case of HD 106906, a resolved scattered light debris disk as well. Gold points are the newly resolved scattered light debris disks presented in this study. As a young moving group, Sco-Cen has a rich population of debris disks with a variety of morphologies and geometries. \textbf{References:} Disk images \citep{esposito2019}, HIP 65426 \citep{chauvin2017}, HD 95086 \citep{rameau2013}, HD 106906 \citep{kalas2015,bailey2014}, Sco-Cen Map \citep{dezeeuw1999}. The references for the discovery papers reporting the first resolved scattered light image of each disk are listed in Table \ref{surveytable}, along with the instrument that first resolved the disk.}
    \label{surveyfigure}
\end{figure*}

The four newly resolved Sco-Cen disks can be combined with the results of related GPI programs to investigate the range of disk structures present in a set of stars with a limited mass range associated with A/F-stars, a common formation environment of an OB Association, and a restricted age range of $\sim$10--15 Myr. The scattered light disk structures discussed in this section will be compared with ALMA results in \S \ref{ALMAcomp}. GPI has resolved one circumstellar disk in Sco-Cen that does not have an A/F host star (HD 129590, see \citealt{matthews2017}), but this disk is not included in our analysis because of its G3 host star. The frequency of resolving disks is beyond the scope of this discovery paper, and is addressed by \cite{esposito2019} in an analysis of the entire GPIES disk survey. Figure \ref{surveyfigure} shows the disk images, revealing the diversity of disks and planets resolved with infrared imaging among 17 A/F-stars in Sco-Cen. The discovery images of the resolved disks and companions were made with several instruments, as noted in Table \ref{surveytable}, but the disk gallery in Figure \ref{surveyfigure} is mainly composed of GPI maps for a more uniform view. Table \ref{surveytable} lists the basic stellar and SED-fit properties along with the source of the resolved scattered light disk discovery and notes from the discovery papers about the morphology and brightness distribution.

Of the 17 systems, 15 have resolved scattered light disks and/or imaged giant planet companions (references in Table \ref{surveytable}), including one system -- HD 106906 -- with both a resolved disk and an imaged planet \citep{kalas2015,bailey2014}. Two of 17 Sco-Cen members have imaged giant planets and no resolved disk in scattered-light infrared imaging, HD 95086 \citep{rameau2013} and HIP 65426 \citep{chauvin2017}. HD 95086 has excess emission based on its SED, while HIP 65426 does not \citep{chen2014}. The majority of the resolved disks have inclinations that are nearly edge-on, however three of the newly resolved disks -- HD 111161, HD 117214, and HD 145560 -- have less inclined geometries which are important for follow-up investigations of the scattering phase function, since lower inclination disks provide access to small scattering angles blocked by the coronagraph in edge-on disks and to portions of the far side of the disk that cannot be disentangled from the front side in an edge-on case. Considering the results of all the resolved systems as summarized in Table \ref{surveytable}, disks that are asymmetric in brightness distribution or morphology appear as common as symmetric structures, highlighting the importance of high angular resolution imaging, since photometry and spectroscopy cannot directly reveal disk structural features.

    \begin{table*}[t]
        \centering
            \begin{tabular}{c S c c c c M B}
                \hline
                \hline
                Name & Spectral Type & $L_{IR}/L_*$ & Instrument & Ref & Disk Type & Morphology & Scattered Light Brightness Distribution\\ \hline
                \multicolumn{8}{c}{Lower Centaurus Crux} \\ \hline
                HD 95086 & A8III & $7.4\times10^{-4}$ & NACO & 1, 3 & Debris & Imaged Planet & N/A \\
                HD 98363 & A8III & $6.4\times10^{-4}$ & GPI & 1, 4 & Debris & Asymmetric \& co-moving companion & Asymmetric \\
                HD 100546 & A0V & --- & NICMOS2 & 5, 16 & Transition & Asymmetric with multiple arms & Asymmetric \\
                HD 106906 & F5V & $4.6\times10^{-4}$ & GPI & 1, 6 & Debris & Asymmetric \& imaged planet & Asymmetric \\
                HD 110058 & A0V & $1.4\times10^{-3}$ & SPHERE & 1, 7 & Debris & Edge-on, wing-tilt asymmetry & Symmetric \\
                HD 111161 & A3III & $5.5\times10^{-4}$ & GPI & 2, 4 & Debris & Inclined ring & Symmetric \\
                HD 111520 & F5V & $6.4\times10^{-4}$ & STIS & 1, 8 & Debris & Edge-on & Asymmetric \\
                HD 114082 & F3V & $3.3\times10^{-3}$ & SPHERE & 1, 9 & Debris & Narrow ring & Asymmetric \\
                HD 115600 & F2IV/V & $1.7\times10^{-3}$ & GPI & 1, 10 & Debris & Ring & Symmetric \\
                HIP 65426 & A2V & 0 & SPHERE & 1, 14 & No disk & Imaged planet & N/A \\
                HD 117214 & F6V & $2.4\times10^{-3}$ & GPI & 1, 11 & Debris & Inclined Ring & Symmetric \\ 
                AK Sco & F5V & --- & SPHERE & 1, 15 & Protoplanetary & Possible gap & Asymmetric \\ \hline
                \multicolumn{8}{c}{Upper Centaurus Lupus} \\ \hline
                HD 131835 & A2IV & $1.5\times10^{-3}$ & GPI & 1, 12 & Debris & Inclined & Asymmetric \\
                HD 143675 & A5IV/V & $4.1\times10^{-4}$ & GPI & 1, 4 & Debris & Edge-on, compact & Symmetric \\
                HD 145560 & F5V & $1.4\times10^{-3}$ & GPI & 1, 4 & Debris & Broad, face-on & Symmetric \\ 
                HD 156623 & A0V & $3.8\times10^{-3}$ & GPI & 2, 11 & Debris & Broad, near face-on, ring & Symmetric \\\hline
                \multicolumn{8}{c}{Upper Scorpius} \\ \hline
                HD 146897 & F2V & $5.3\times10^{-3}$ & HiCIAO & 1, 13 & Debris & Edge-on, stellocentric offset & Symmetric \\
                \hline

            \end{tabular}
        \caption{Spectral and disk properties of resolved scattered light circumstellar disks and planets in Sco-Cen. \textbf{References:} 1. \cite{chen2014} 2. \cite{mcdonald2012} 3. \cite{rameau2013}, 4. this work 5. \cite{currie2015a} 6. \cite{kalas2015} 7. \cite{kasper2015} 8. \cite{draper2016} 9. \cite{wahhaj2016} 10. \cite{currie2015b} 11. \cite{esposito2019} 12. \cite{hung2015} 13. \cite{thalmann2013} 14. \cite{chauvin2017} 15. \cite{janson2016}. 16. \cite{augereau2001}}
        \label{surveytable}
    \end{table*}

\subsection{Comparisons of Infrared Scattered Light Disk Images with ALMA Millimeter Maps} \label{ALMAcomp}
The GPI near-IR scattered light images that preferentially probe the population of smaller micron-sized dust grains can be compared with ALMA millimeter maps of the dust continuum emission that is sensitive to the larger mm-sized dust particles in the disk. Of the four newly resolved targets in this study, all were observed with ALMA, and the main results from the continuum and line ALMA data are given in Table \ref{ALMAtable}. None of the four targets have gas disk detections in the CO(2-1) line \citep{lieman-sifry2016,moor2017}. The 1.24~mm continuum fluxes range from a non-detection for the most compact scattered light disk around HD 143675 to the 1850 $\mu$Jy strong detection around the broad, near face-on HD 145560 disk \citep{moor2017,lieman-sifry2016}. Two of the newly resolved scattered light disks -- HD 98363 and HD 111161 -- have faint and unresolved ALMA 1.24mm detections.  

For the HD 145560 disk, the ALMA continuum emission is spatially resolved along both axes, which is broadly consistent with the model estimation of an outer radius beyond the extent of the GPI field-of-view \citep{esposito2019}. The ALMA-based estimates of $R_{\mathrm{in}}$ of $56^{+11}_{-9}$ and $R_{\mathrm{out}}$ of $126^{+20}_{-30}$ from a morphological modeling analysis \citep{lieman-sifry2016} are indicated on the GPI disk image in Figure \ref{modelSED}. The inner disk radius estimate is consistent with the results from GPI data modeling reported in Table \ref{modeloutputs} \citep{esposito2019}. The outer disk radii estimates are not consistent with each other, however each approach to determining $R_{\mathrm{out}}$ has significant limitations. The $R_{\mathrm{out}}$ from modeling the GPI scattered light imaging is poorly constrained due to the low surface brightness of the data at larger radial separations \citep{esposito2019}, while the $R_{\mathrm{out}}$ estimated from ALMA data was based on a fixed surface density power law index, a parameter which is degenerate with the outer radius \citep{lieman-sifry2016}.  

\begin{figure*}
    \centering
    \includegraphics[width=0.8\textwidth]{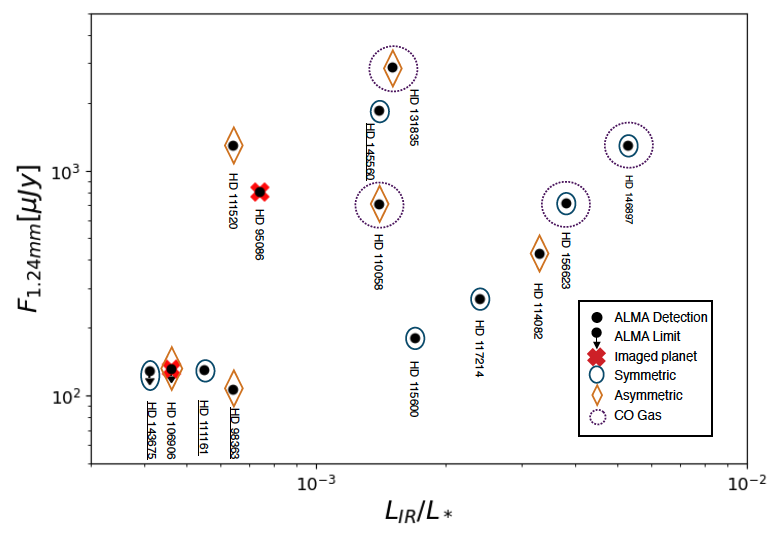}
    \caption{ALMA disk detections (points) and upper limits (point with arrow) as a function of IR excess, indicating systems with scattered light disks that are symmetric (oval) or asymmetric (diamond). Stars with planets have a red 'x' and systems with CO(2-1) detections have large dashed circles. Most targets show a systematic positive trend of increasing $F_{1.24mm}$ with higher $L_{IR}/L_*$, though five of the disks have high ALMA fluxes despite lower IR excesses. Possible explanations for this could be the presence of lower surface brightness extended disks or halos. The systems with imaged planets (including one star not on the plot due to a lack of an excess) do not have symmetric scattered light disks or CO(2-1) gas detections. \textbf{References:} 1. \cite{lieman-sifry2016}, 2. \cite{moor2017}, 3. \cite{su2017}}
    \label{ALMAfigure}
\end{figure*}

The ALMA results on the larger set of early-type Sco-Cen members with resolved infrared disks and imaged planets are also compiled in Table \ref{ALMAtable}; 15 of the 17 systems in Table \ref{surveytable} have ALMA measurements. The HD 145560 disk is the second brightest in the millimeter of these debris disks around early-type A/F-star members with resolved scattered light disks or imaged planets. Excluding the systems AK Sco and HD 100546 with disks at an earlier evolutionary state and the HIP 65426 system with no detectable infrared excess, the ALMA 1.24~mm debris disk fluxes are plotted as a function of the IR excess in Figure \ref{ALMAfigure}. The IR excesses are from the cooler second blackbody fit in the \cite{chen2014} analysis or the single fit from the \cite{mcdonald2012} study. The data show a large amount of scatter in Figure \ref{ALMAfigure}, particularly among the lower IR excess level systems which includes a group of 5 targets with high millimeter fluxes despite low IR excess levels.

The combination of the scattered-light image and millimeter maps can be used to consider possible explanations for these high $F_{\mathrm{1.24mm}}$/low IR excess systems which have colder dust. A higher frequency of asymmetry in these disks is not caused by a possible detection bias associated with larger disks having more easily identifiable asymmetries, since these systems are not the largest disks in the Sco-Cen sample. Although the sample is small, the resolved scattered-light disks in this category are typically asymmetric (3 of 4 systems) compared to disks with systematically increasing 1.2~mm flux as a function of infrared excess (only 2 of 7 are asymmetric). The HD 95086 and HD 145560 systems present cases in which there is clear evidence or a strong indication of a very extended disk with a central clearing or low dust density region. For HD 95086, the central clearing is large enough to make the disk undetected in scattered light within the limited GPI field-of-view while a large ring is imaged in wider field ALMA maps \citep{su2017}. Higher sensitivity wider field infrared imaging of these systems may detect lower surface brightness extended disks or halos.

Figure \ref{ALMAfigure} also highlights the four debris disks with CO gas detections; the ALMA bandwidth covered the CO(2-1) line for each of the objects in Table \ref{ALMAtable}. The systems with gas disks all have high 1.24~mm continuum fluxes, though not every high $F_{1.24mm}$ disk has a corresponding CO detection. The debris disks with gas exhibit a diversity of structures in the scattered-light images and span the full range of $L_{IR}/L_*$ values. Of the three Sco-Cen members with imaged giant planets, two have been observed with ALMA and neither retains a CO gas disk. $\beta$ Pictoris (not in Sco-Cen) is a contrary example of a system that contains an imaged planet, debris disk, CO gas emission \citep{matra2017}, and CI gas emission \citep{cataldi2018}. The three Sco-Cen A/F-stars with imaged giant planets include one system with no excess (HIP 65426) and two (HD 95086 and HD 106906) with IR excesses in the lower half of the 16 debris disk systems listed in Table \ref{surveytable}. Only one imaged planetary system, HD 106906, has a resolved scattered-light disk, and the HD 106906 disk reveals a very asymmetric structure \citep{kalas2015}. Although the total number of imaged planetary systems is limited, Sco-Cen contains the largest number of such systems in any one stellar population. Sco-Cen also has a large population of resolved debris disks \citep{esposito2019} and has an age associated with the peak of IR excess emission \citep{wyatt2008}.

\begin{table*}[t]
    \centering
    \begin{tabular}{c c c c c c c c c S c}
    \hline
    \hline
        Name & HIP & $L_{IR}/L_*$ & Disk Type & Symmetric? & F [$\mu$Jy] & $\lambda$ [mm] & Beam Size & Resolved? & CO Det [mJy km s$^{-1}$] & Ref \\ \hline
        \multicolumn{11}{c}{Lower Centaurus Crux} \\ \hline
        HD 95086 & 53524 & $7.4\times10^{-4}$ & Debris & N/A & 810 & 1.3 & 1.22 $\times$ 1.03 & 2 axes & N & 1 \\
        HD 98363 & 55188 & $6.4\times10^{-4}$ & Debris & N & 107 & --- & 0.7$\times$0.82 & 2 axes & N & 2 \\
        HD 100546 & 56379 & --- & Transition & N & --- & --- & --- & --- & --- & --- \\
        HD 106906 & 59960 & $4.6\times10^{-4}$ & Debris & N & $<$132 & 1.24 & --- & N/A & N & 3 \\
        HD 110058 & 61782 & $1.4\times10^{-3}$ & Debris & Y & 710 & 1.24 & 1.36$\times$0.83 & 1 axis & 5.5 & 3\\
        HD 111161 & 62482 & $5.5\times10^{-4}$ & Debris & Y & 130 & 1.24 & 1.3$\times$1.0 & unresolved & N & 3 \\
        HD 111520 & 62657 & $6.4\times10^{-4}$ & Debris & N & 1290 & 1.24 & 1.37$\times$0.83 & 1 axis & N & 3 \\
        HD 114082 & 64184 & $3.3\times10^{-3}$ & Debris & N & 430 & 1.24 & 1.32$\times$0.89 & unresolved & N & 3 \\
        HD 115600 & 64995 & $1.7\times10^{-3}$ & Debris & Y & 180 & 1.24 & 1.32$\times$0.88 & unresolved & N & 3 \\
        HIP 65426 & 65426 & 0 & No Disk & N/A & --- & --- & --- & --- & --- & --- \\
        HD 117214 & 65875 & $2.4\times10^{-3}$ & Debris & Y & 270 & 1.24 & 1.32$\times$0.86 & unresolved & N & 3 \\
        AK Sco & 82747 & --- & Protoplanetary & N & 35930 & 1.24 & 1.22$\times$0.76 & 2 axes & 10.5 & 3 \\ \hline
        \multicolumn{11}{c}{Upper Centaurus Lupus} \\ \hline
        HD 131835 & 73145 & $1.5\times10^{-3}$ & Debris & N & 2900 & 1.24 & 1.36$\times$1.16 & 2 axes & 22.5 & 3 \\
        HD 143675 & 78641 & $4.1\times10^{-4}$ & Debris & Y & $<$129 & 1.24 & 0.48$\times$0.64 & N/A & N & 1 \\
        HD 145560 & 79516 & $1.4\times10^{-3}$ & Debris & Y & 1850 & 1.24 & 1.25$\times$0.82 & 2 axes & N & 3 \\
        HD 156623 & 84881 & $3.8\times10^{-3}$ & Debris & Y & 720 & 1.24 & 1.25$\times$0.82 & 1 axis & 32.3 & 3 \\ \hline
        \multicolumn{11}{c}{Upper Scorpius} \\ \hline
        HD 146897 & 79977 & $5.3\times10^{-3}$ & Debris & Y & 1300 & 1.24 & 1.05$\times$0.67 & 1 axis & 4.1 & 3 \\
         \hline
    \end{tabular}
    \caption{Sco-Cen A and F type stars with known disks and/or companions as observed with ALMA. In almost all cases aside from HD 106906 and HD 143675, 1.24 mm flux resolved and unresolved detections were achieved. CO detections were found with some of the disks. \textbf{References:} 1. \cite{su2017} 2. \cite{moor2017} 3. \cite{lieman-sifry2016}}
    \label{ALMAtable}
\end{table*}

\section{Summary} \label{conclusion}
We have spatially resolved four Sco-Cen debris disks for the first time in scattered light using the Gemini Planet Imager. The four debris disk systems were targeted by GPI due to their high IR excess emission, with three of their SEDs best fit by a two-temperature model. The four debris disks---HD 98363, HD 111161, HD 143675, and HD 145560---were all resolved in polarized intensity light. HD 143675 was also resolved in total intensity light using the spectral mode of GPI. HD 111161 and HD 145560 show debris disks that are diffuse and moderately inclined. HD 143675 presents a debris disk that is quite compact with a highly inclined, edge-on geometry. Preliminary results of the HD 98363 disk show a highly inclined and diffuse structure that is also asymmetric in its brightness distribution. Surface brightness profiles measured for HD 111161, HD 143675, and HD 145560 show a symmetric brightness distribution, while the HD 98363 map shows the NE side of the disk to be tentatively brighter and more radially extended compared to the SW side of the disk. The disk images are compared with the results of SED-fitting models \citep{jang-condell2015,mcdonald2012} and radiative transfer models \citep{esposito2019} to investigate the architectures of the disk systems. The four debris disks were also observed with ALMA \citep{lieman-sifry2016,moor2017,su2017} and the results are compared with GPI scattered light imaging. The best fitting model for the debris disk around HD 145560 suggests extended emission beyond the field-of-view for GPI, which is also consistent with its ALMA millimeter map. In the case of HD 143675, the most compact system, no 1.24mm flux emission was detected, while HD 98363 and HD 111161 both had faint and unresolved detections. None of the disks have detectable gas emission \citep{moor2017,lieman-sifry2016}. 

The debris disk around HD 98363 is a unique case with a wide, $\sim$7000 AU binary companion Wray 15-788, with its own circumstellar disk classified as a transitional disk at an earlier evolutionary state \citep{bohn2019}. In addition to being in different stages of circumstellar disk evolution, the inclinations of both disks are misaligned ($\Delta i \sim 60\degr$). Similarities can be made between the HD 98363 system and the HD 106906 system as they both have similar morphological properties. The HD 98363/Wray15-788 system presents the ideal case for future studies of binary-disk interactions. It is unclear whether or not the asymmetry of HD 98363 and/or the two-belt structure of Wray 15-788 were caused by mutual dynamical perturbations. It is also possible that the asymmetry in HD 98363 could be caused by a much closer, planetary mass companion within the system, although no giant planet was detected in the GPI data. 

Combining the newly-resolved debris disk systems with other examples reveals Scorpius-Centaurus as the site of a population of circumstellar disks with a range of disk structures. The full set of GPI-resolved Sco-Cen scattered light disks around early-type stars includes one protoplanetary, one transitional, and 14 debris disks with morphologies that vary in inclination, asymmetry, vertical structure, and size. By comparing ALMA millimeter maps to GPI-resolved scattered light images of Sco-Cen debris disks, a diverse combination of properties are observed without a single unifying pattern, however stars with low IR excess and high mm flux typically exhibit asymmetric scattered light disks, while stars with an IR excess that scales with mm flux typically exhibit symmetric scattered light disks. If the disk asymmetry is caused by dynamical interactions with an undetected companion, then the higher mm fluxes may be analogous to the high fluxes measured for circumbinary disks in the younger Taurus region \citep{harris2012}. For the specific case of HD 95086, a planetary companion has been imaged and it may have cleared a substantial portion of the inner disk material, causing a lower IR excess, and resulted in an extended outer dust disk with higher ALMA flux. CO gas detections are also present around disks with high 1.24mm flux emission, but gas is not detected around every disk with high 1.24mm flux emission or any of the targets or Sco-Cen members with imaged planets that were observed with ALMA. 

Advanced direct imaging instruments such as GPI or SPHERE have revealed fine disk structure which cannot be inferred from spectral energy distributions or millimeter maps with a coarse beam. The GPI scattered light maps can be used to motivate future studies of these systems at a range of spatial and spectral resolutions across multiple wavelengths.

\section*{Acknowledgements}
This work is based on observations obtained at the Gemini Observatory, which is operated by the Association of Universities for Research in Astronomy, Inc., under a cooperative agreement with the NSF on behalf of the Gemini partnership: the National Science Foundation (United States), the National Research Council (Canada), CONICYT (Chile), Ministerio de Ciencia, Tecnolog\'ia e Innovaci\'on Productiva (Argentina), and Minist\'erio da Ci\^encia, Tecnologia e Inova\c c\~ao (Brazil). This work has made use of data from the European Space Agency (ESA) mission {\it Gaia} (\url{https://www.cosmos.esa.int/gaia}), processed by the {\it Gaia} Data Processing and Analysis Consortium (DPAC, \url{https://www.cosmos.esa.int/web/gaia/dpac/consortium}). Funding for the DPAC has been provided by national institutions, in particular the institutions participating in the {\it Gaia} Multilateral Agreement. This research has made use of the SIMBAD and VizieR databases, operated at CDS, Strasbourg, France.

Supported by NSF grants AST-1411868 (E.L.N., K.B.F., B.M., and J.P.), AST-141378 (G.D.), and AST-1518332 (R.D.R., J.J.W., T.M.E., J.R.G., P.K., G.D.). Supported by NASA grants NNX14AJ80G (E.L.N., S.C.B., B.M., F.M., and M.P.), NNX15AC89G and NNX15AD95G/NExSS (B.M., J.E.W., T.M.E., R.J.D.R., G.D., J.R.G., P.K.), NN15AB52l (D.S.), and NNX16AD44G (K.M.M.). J.R., R.D. and D.L. acknowledge support from the Fonds de Recherche du Qu\`ebec. J.R.M.'s work was performed in part under contract with the California Institute of Technology (Caltech)/Jet Propulsion Laboratory (JPL) funded by NASA through the Sagan Fellowship Program executed by the NASA Exoplanet Science Institute. M.M.B. and J.M. were supported by NASA through Hubble Fellowship grants \#51378.01-A and HST-HF2-51414.001, respectively, and I.C. through Hubble Fellowship grant HST-HF2-51405.001-A, awarded by the Space Telescope Science Institute, which is operated by the Association of Universities for Research in Astronomy, Inc., for NASA, under contract NAS5-26555. K.W.D. is supported by an NRAO Student Observing Support Award SOSPA3-007. J.J.W. is supported by the Heising-Simons Foundation 51 Pegasi b postdoctoral fellowship. This work benefited from NASA's Nexus for Exoplanet System Science (NExSS) research coordination network sponsored by NASA's Science Mission Directorate. Portions of this work were also performed under the auspices of the U.S. Department of Energy by Lawrence Livermore National Laboratory under Contract DE-AC52-07NA27344.

\software{Gemini Planet Imager Data Pipeline (\citealt{perrin2014,perrin2016}, \url{http://ascl.net/1411.018}), pyKLIP (\citealt{wang2015}, \url{http://ascl.net/1506.001}), numpy, scipy, Astropy \citep{astropy2018}, matplotlib \citep{matplotlib2007, matplotlib_v2.0.2}, iPython \citep{ipython2007}, emcee (\citealt{foreman-mackey2013}, \url{http://ascl.net/1303.002}), corner (\citealt{foreman-mackey2017}, \url{http://ascl.net/1702.002})}.

\facilities{Gemini:South}
\bibliographystyle{aasjournal}
\bibliography{export-bibtex}

\end{document}